\documentclass[
 aps,
 prl,
 reprint,
 superscriptaddress,
 amsmath,
 amssymb,
 floatfix
]{revtex4-2}

\usepackage{graphicx}
\usepackage{bm}
\usepackage{dcolumn}
\usepackage{xcolor}
\usepackage[colorlinks=true,citecolor=blue,linkcolor=blue,urlcolor=blue]{hyperref}

\graphicspath{{figures/}}
\newcommand{\angstrom}{\mathring{\mathrm{A}}}

\begin{document}

\title{Moir\'e Phonon Condensation in Magic-Angle Twisted Bilayer Graphene}

\author{Zhanghao Zhouyin}
\affiliation{Department of Physics, McGill University, Montreal, Quebec, Canada H3A2T8}
\email{yinzhanghao.zhou@mail.mcgill.ca}

\author{Jyun-Jie Jiang}
\affiliation{Department of Physics, McGill University, Montreal, Quebec, Canada H3A2T8}

\author{Xianghua Kong}
\affiliation{College of Physics and Optoelectronic Engineering, Shenzhen University, Shenzhen,
518060, China}

\author{Hong Guo}
\affiliation{Department of Physics, McGill University, Montreal, Quebec, Canada H3A2T8}

\date{\today}

\begin{abstract}
Twisted bilayer graphene reconstructs from weak breathing corrugation to large common bending near the magic angle, but the origin of this collective crossover has remained unclear. Here we show that the crossover is a soft-mode condensation of layer-symmetric $A_1$ moir\'e flexural phonons: these modes soften on the breathing branch, lose stiffness near the magic angle, and freeze into the bending morphology. We call this mechanism Moir\'e Phonon Condensation (MPC). At $\theta=1.08^\circ$, it is extremely surprising that displacements of all 11164 atoms in the moir\'e supercell, with a maximum atomic position shift of $2.30\,\angstrom$, is captured by only two $A_1$ phonon modes at more than $99.5\%$ spectral weight. A first-harmonic continuum theory identifies a dimensionless control parameter of the phenomenon, showing that as the twist approaches the magic angle, the growing moir\'e length scale amplifies a smooth stress-bending competition until the flexural stiffness changes sign. Mode-resolved tight-binding calculations further show that the condensed phonon coordinates are electronically active. This work identifies MPC as a twist-controlled structural order parameter for moir\'e reconstruction. 
\end{abstract}

\maketitle

Bilayer moir\'e graphene has emerged as a highly tunable platform for engineering quantum materials. Near a magic angle, the moir\'e potential produces narrow electronic bands that amplify electron-electron interaction and enable correlated insulating phases, superconductivity, orbital magnetism, and other emergent quantum phenomena. As a controlling factor of the moir\'e potential, lattice reconstruction is of central importance: relaxation expands energetically favorable AB/BA stacking, suppresses AA stacking, and produces out-of-plane corrugation that changes interlayer tunneling, flat-band gaps, and local electronic texture \cite{Bistritzer2011,Dai2016,NamKoshino2017,Carr2018,Gargiulo2018,Yoo2019,Lucignano2019,Cantele2020,Rakib2022,Li2024}. A particularly striking feature is the crossover between two corrugation patterns: weak breathing corrugation, dominated by modulation of the interlayer distance, and large common bending of the two graphene sheets \cite{Dai2016,Rakib2022,jain2017structure}. Although this breathing-to-bending crossover has been established, the dynamical lattice coordinate that drives it has remained undefined.

Here, we identify this missing coordinate and show that magic-angle twisted bilayer graphene realizes an exceptionally clean moir\'e-scale soft-mode condensation.  As the twist angle $\theta$ is reduced, layer-symmetric $A_1$ moir\'e flexural modes soften on the breathing branch, become unstable near the magic angle, and condense into the bending morphology. 
At $\theta=1.08^\circ$, the displacement from the breathing saddle to the bending minimum is captured by just two $A_1$ phonon coordinates with more than $99.5\%$ spectral weight, despite involving all 11164 atoms and a maximum atomic position shift of $2.30\,\angstrom$. We name this twist-angle-driven soft-mode transition \emph{Moir\'e Phonon Condensation} (MPC): a moir\'e-scale phonon coordinate loses stiffness and freezes, acting as a lattice order parameter for the reconstruction, as in the soft mode theory~\cite{cochran1960crystal,cowley1980structural,dove1997theory,orihara2002observation}.

Previous studies have largely treated moir\'e reconstruction as an endpoint structural-relaxation problem, rather than as a soft-mode instability.
For a given twist angle, atomistic and continuum models minimize the total energy, obtain the relaxed atomic displacement, and use the resulting geometry to analyze electronic tunnelling, flat bands, energy gaps, and real-space electronic textures~\cite{uchida2014atomic,wijk2015relaxation,Dai2016,jain2017structure}. 
In parallel, studies of moir\'e phonons have established that twisted bilayer lattices possess low-energy vibrational modes on the very large moir\'e scale that couple to the electronic structure~\cite{cocemasov2013phonons,koshino2019moire,Angeli2019,KoshinoNam2020,Liu2022}. 
However, these two views have remained largely separate. What has been missing is the direct identification of the moiré phonon coordinates that generate the reconstructed morphology itself.

\begin{figure}[t!]
\centering
\includegraphics[width=\columnwidth]{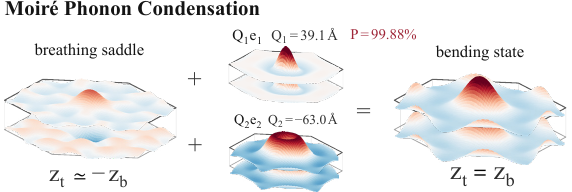}
\caption{\label{fig:overview}
Overview of Moir\'e Phonon Condensation. The breathing-to-bending reconstruction at \(\theta=1.08^\circ\) is shown as a finite condensation of two unstable layer-symmetric \(A_1\) modes. From left to right, it shows the breathing saddle, the two-mode $A_1^{(1)}$ and $A_1^{(2)}$ landscape, and the bending endpoint in the same moir\'e cell; The breathing saddle is layer anti-symmetric \(z_{\rm t}\simeq-z_{\rm b}\), while the bending state has \(z_{\rm t}\simeq z_{\rm b}\). The two-mode projection captures \(99.881\%\) of the REBO+KC endpoint displacement and \(99.539\%\) in the MLFF calculation.}
\end{figure}

The MPC mechanism is naturally connected to the soft-mode theory of displacive structural phase transitions~\cite{cochran1960crystal,cowley1980structural,dove1997theory,orihara2002observation}. 
In that theory, the crucial object is not only the final distorted structure, but the phonon mode whose squared frequency is driven to zero by a control parameter. 
Once the stiffness becomes negative, anharmonicity selects a finite amplitude of the soft eigenvector, which becomes the structural order parameter. 
For a moir\'e supercell, such a reduction is far from guaranteed: an Angstrom-scale reconstruction involving thousands of atoms could in principle spread over many modes through anharmonic couplings, strain, and secondary distortions. 
The near-complete confinement of the breathing-to-bending pathway to two $A_1$ modes therefore reveals an unusually clean soft-mode mechanism. 
Below, we further derive a first-harmonic continuum theory showing that the condensation is driven by the growing moir\'e length scale, which amplifies an otherwise smooth stress--bending competition through the geometric factor $g^{-2}$, producing a definite twist-angle softening trend.

\begin{figure*}[t!]
\centering
\includegraphics[width=0.84\textwidth]{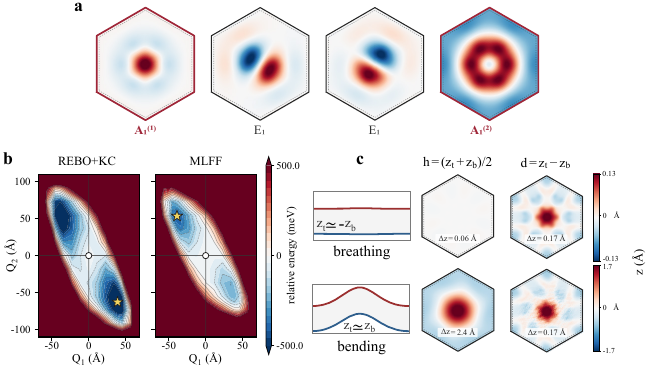}
\caption{\label{fig:atomic}
Atomistic signature of Moir\'e Phonon Condensation. Panel (a) shows the layer-symmetric height patterns of the unstable modes; mode $A_1^{(1)}$ and $A_1^{(2)}$ are highlighted because they form the condensing subspace. Panel (b) shows the energy landscape of the magic angle moir\'e graphene system along the two collapsed phonon branches as $\mathbf{R}=\mathbf{R}_{\text{breathing}}'+Q_1\mathbf{e}_1+Q_2\mathbf{e}_2$. Here, the initial breathing structure is corrected by the $<0.5\%$ of residual amplitude, denoted as $\mathbf{R}_{\text{breathing}}'$. Panel (c) separates the relaxed structures into layer-antisymmetric spacing modulation \(d=z_t-z_b\) and layer-symmetric mid-surface height \(h=(z_t+z_b)/2\). The bending endpoint creates only small vortex in \(d({\bf r})\) and shares similar texture as the breathing phase, while the \(h({\bf r})\) grows from a weak residual corrugation (\(0.06\,\angstrom\)) to a large common bending (\(2.4\,\angstrom\)).}
\end{figure*}

Fig.~\ref{fig:overview} summarizes the MPC. The displacement from breathing to bending at \(\theta=1.08^\circ\) can be represented by only two phonon eigenmode basis vectors,
\begin{equation}
{\bf R}_{\rm bending}-{\bf R}_{\rm breathing}\simeq Q_1{\bf e}_1+Q_2{\bf e}_2,
\end{equation}
with essentially no other component in the full \((3N-6)\) dimensional vibrational space where $N$ is the number of atoms in the supercell. Here $Q_1,Q_2$ are the projected mode amplitudes, $\mathbf{e}_1,\mathbf{e}_2$ are the normalized eigenmodes. The displacement field is overwhelmingly layer-symmetric, as characterized by the height profiles: weak breathing has \(z_{\rm t}\simeq -z_{\rm b}\), whereas the condensed state has \(z_{\rm t}\simeq z_{\rm b}\).

The $\mathbf{R}_{\rm breathing}$ is the high-symmetry stationary branch obtained by unconstrained relaxation from a D6-symmetric commensurate initial structure. At large twisting angles, this branch is a true local energy minimum; near the magic angle, the same branch becomes a saddle, as diagnosed by the Hessian. No symmetry constraints are imposed during structural relaxation. The breathing structure creates a vortex texture that is layer-antisymmetric (see left plot in Fig.~\ref{fig:overview}). We then compute the \(\Gamma\)-point force-constant matrix and identify the unstable normal modes as in Fig.~\ref{fig:atomic}(a). The four unstable modes contain two separate $A_1$ modes and two degenerate $E_1$-like modes. We then relax structures kicked along the unstable subspace and calculate the endpoint displacements:
\begin{equation}
\Delta {\bf R}={\bf R}_{\rm bending}-{\bf R}_{\rm breathing},
\end{equation}
where \({\bf R}_{\rm breathing}\) is the weakly corrugated breathing structure and \({\bf R}_{\rm bending}\) is the relaxed bending endpoint. The same bending basin can be stably reached from all tested initial directions in the unstable subspace, indicating that the endpoint is selected by the energy landscape rather than by a fine-tuned initial condition. We project the displacement onto the phonon eigenmode basis \(\{{\bf e}_i\}\). The projection weight in the two selected $A_1^{(1)}$ and $A_1^{(2)}$ modes is
\begin{equation}
P_{1,2}=
\frac{
|\langle {\bf e}_1|\Delta{\bf R}\rangle|^2+
|\langle {\bf e}_2|\Delta{\bf R}\rangle|^2
}{
\|\Delta{\bf R}\|^2
}.
\end{equation}

There is no kinematic reason for a finite relaxed displacement of more than two Angstroms to remain in a two-dimensional tangent plane of the saddle. But surprisingly, it does. At \(\theta=1.08^\circ\), the two unstable flexural modes $A_1^{(1)}$ and $A_1^{(2)}$ account for \(P_{1,2}=99.881\%\) of the breathing-to-bending displacement calculated by the second-generation REBO intralayer potential with a Kolmogorov-Crespi interlayer registry potential (REBO+KC) \cite{krongchon2023registry, brenner2002second}, and \(99.539\%\) by the machine-learning force field (MLFF) model \cite{wang2018deepmd}. The residual weight is below $0.5\%$ in both force field descriptions. This is the operational signature of MPC: the endpoint is selectively confined in a small unstable phonon subspace instead of a generic path through the \((3N-6)\) dimensional configuration space. Details about the force field and calculation settings are available in Section III of the supplemental materials.



The two-mode energy landscape provides the geometric counterpart of the MPC. As in Fig.~\ref{fig:atomic}(b), we use the dominant $A_1^{(1)}$ and $A_1^{(2)}$ modes to form the normal coordinates $Q_1$ and $Q_2$, then compute the energy of the structure
\begin{equation}
    \mathbf{R}(Q_1,Q_2)=\mathbf{R'}_{\text{breathing}}+Q_1\mathbf{e_1}+Q_2\mathbf{e_2}.
\end{equation}
Here $\mathbf{R'}_{\text{breathing}}$ is the breathing reference corrected by residual modes, which is located at the centre flat saddle region. Along the normal coordinates $Q_1$ and $Q_2$, the energy decreases and is stabilised at finite amplitude, forming the bending basin. The projected endpoint lies exactly at the basin minimum, showing that the two unstable phonons are not merely the initial escape directions from the saddle but remain the dominant coordinates during the whole structural reconstruction.

The real-space character of the condensed phonons inspires a simple theory of such a low-dimensional collapse. Let
\begin{equation}
d({\bf r})=z_t({\bf r})-z_b({\bf r}),\qquad
h({\bf r})=\frac{z_t({\bf r})+z_b({\bf r})}{2}.
\end{equation}
Here \(d({\bf r})\) measures the layer-antisymmetric corrugation, or equivalently the local modulation of the interlayer spacing, while \(h({\bf r})\) is the height of the bilayer mid-surface. Fig.~\ref{fig:atomic}(c) gives a clear picture: for \(d=z_t-z_b\), both the breathing saddle and the bending endpoint contain a similar separation pattern of AA regions and AB/BA domains. The bending only creates a slight spiral, and the amplitude increases slightly from \(0.165\,\angstrom\) to \(0.174\,\angstrom\), which can be neglected compared to the change of the \(2.4\,\angstrom\) in the layer symmetric field \(h({\bf r})\). Therefore, the structure transition can be simplified as layer-symmetric movement, which motivates a one-field theory of the instability below. The same type of low-dimensional soft-mode collapse is not unique to bilayer graphene. As summarized in Section I.3 of the Supplemental Material, in twisted hBN bilayers, the breathing-to-bending displacement is dominated by three \(A_1\) moiré flexural modes that account for more than 98.5$\%$ of the total structure reconstruction.

\begin{figure*}[t!]
\centering
\includegraphics[width=0.75\textwidth]{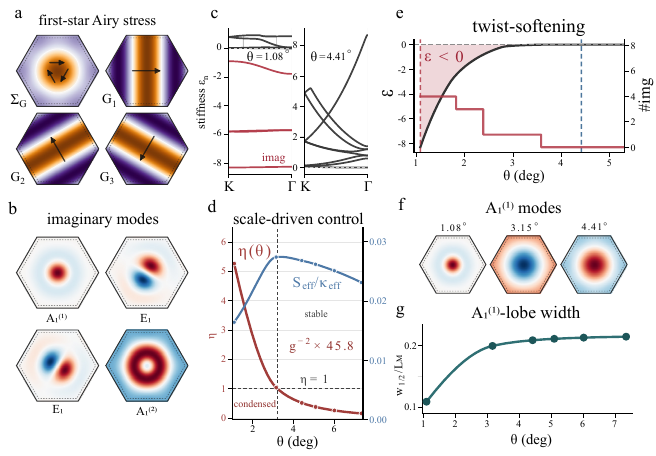}
\caption{\label{fig:theory}
First-harmonic continuum theory of Moir\'e Phonon Condensation. The layer-symmetric height field \(w({\bf r})\) is governed by a bending term and an Airy-stress term generated by the breathing-relaxed moir\'e texture as in Panel (a). Panel (b) shows unstable \(\Gamma\)-point modes at \(1.08^\circ\) twisted graphene bilayer, predicted by the first-harmonic continuum elasticity theory model. Panel (c) is the predicted phonon band structure on two representative twisted angles. Panel (d) shows the extracted control parameter \(\eta=S_{\rm eff}/(\kappa_{\rm eff}g^2)\), it decreases from 5.27 at \(1.08^\circ\) to 0.52 at \(4.41^\circ\), while the factor $S_{\rm eff}/\kappa_{\rm eff}$ varies smoothly. Panel (e) also shows the reduction of imaginary mode numbers and the increase of the minimum mode eigenvalue. Panels (f) and (g) show the \(A_1^{(1)}\) modes and the monopolar area's width across angles.}
\end{figure*}

The atomistic collapse suggests that the instability can be reduced to a continuum theory for the layer-symmetric height fluctuation above the breathing reference,
\begin{equation}
w({\bf r},t)=\frac{\delta z_t({\bf r},t)+\delta z_b({\bf r},t)}{2},
\end{equation}
while the layer-antisymmetric fluctuation is small during structure reconstruction. This keeps the interlayer distance \(d({\bf r})\) fixed and isolates the common bending coordinate. The symmetry-allowed quadratic energy is:
\begin{equation}
\label{eq:energy}
E[w]=
\frac12\int d^2r\,
\left[
\kappa_{\rm eff}(\nabla^2 w)^2+
\Sigma^{\rm eff}_{ij}({\bf r};\theta)
\partial_i w\,\partial_j w
\right].
\end{equation}
The first term penalises curvature of the membrane, the second term is the geometric coupling between height gradients and the effective in-plane stress already stored in the breathing-relaxed moir\'e texture.

For a mechanically equilibrated two-dimensional state, \(\partial_i\Sigma^{\rm eff}_{ij}=0\), and the periodic stress can be represented by an Airy function. As visualized in Fig.~\ref{fig:theory}(a), we keep only the first moir\'e star,
\begin{equation}
\label{eq:Airy}
\Phi({\bf r})=\Phi_1\sum_{m=1}^{3}\cos({\bf G}_m\cdot{\bf r}),
\end{equation}
where \(\mathbf{G}_m\) is the reciprocal vector of moir\'e lattice, and \(g=|{\bf G}_m|\). This first-harmonic truncation is the minimal form that preserves the moir\'e \(D_6\) symmetry; it also reduces the tensor stress field to a single amplitude $\Phi_1$. Taking the stress tensor from the Airy function and transforming to Fourier space, the first-harmonic components can be written as, with \({\bf q}_m=\hat z\times{\bf G}_m\):
\begin{equation}
\Sigma_{ij}^{\rm first}(\pm{\bf G}_m)
=-\frac{S_{\rm eff}}{2g^2}q_{m,i}q_{m,j},
\qquad
S_{\rm eff}\equiv\Phi_1g^2 .
\end{equation}
Here \(S_{\rm eff}\) is the amplitude of the first Airy-stress harmonic, expressed in stress units. Substituting this field into Eq.~\eqref{eq:energy} and expanding \(w({\bf r})=\sum_{\bf G}w_{\bf k}(\bf G)e^{i({\bf k}+{\bf G})\cdot{\bf r}}\) gives a dimensionless eigenproblem after factoring out \(\kappa_{\rm eff}g^4\):
\begin{equation}
\begin{aligned}
\sum_{{\bf G}'} h^{\rm first}_{{\bf G}{\bf G}'}({\bf k})w_{{\bf k}}({\bf G}')
&=\varepsilon_n({\bf k})w_{\bf k}({\bf G}),\\
\omega_n^2({\bf k})=
\frac{\kappa_{\rm eff}g^4}{\rho_{\rm eff}}\varepsilon_n({\bf k}).
\end{aligned}
\label{eq:first_star_eigen}
\end{equation}
Negative \(\varepsilon_n\) therefore corresponds directly to an imaginary flexural phonon. The dimensionless matrix is
\begin{equation}
\label{eq:first_star_h}
\begin{aligned}
h^{\rm first}_{{\bf G}{\bf G}'}=
|\widetilde{\bf K}|^4\delta_{{\bf G}{\bf G}'}
&-
\frac{\eta}{2}
\sum_{m=1}^{3}
\left(\widetilde{\bf q}_m\cdot\widetilde{\bf K}\right)
\left(\widetilde{\bf q}_m\cdot\widetilde{\bf K}'\right)
\\
&\times
\left[
\delta_{{\bf G}-{\bf G}',{\bf G}_m}
+\delta_{{\bf G}-{\bf G}',-{\bf G}_m}
\right].
\end{aligned}
\end{equation}
where \(\widetilde{\bf K}=({\bf k}+{\bf G})/g\), \(\widetilde{\bf K}'=({\bf k}+{\bf G}')/g\), and \(\widetilde{\bf q}_m={\bf q}_m/g\). The single control parameter in Eq. (\ref{eq:first_star_h}) is
\begin{equation}
\eta(\theta)=
\frac{S_{\rm eff}(\theta)}
{\kappa_{\rm eff}(\theta)g^2(\theta)} .
\end{equation}
This ratio separates the ingredients of the instability. \(S_{\rm eff}\) measures the first-harmonic internal stress left by relaxation; for the soft flexural pattern it contributes negative stiffness. \(\kappa_{\rm eff}\) is the corresponding bending penalty, and \(g^2\) is the square of the reciprocal vector length of the moir\'e lattice. As discussed in Section II of the supplementary, \(S_{\rm eff}\) and \(\kappa_{\rm eff}\) are extracted by finite-difference calculations around the relaxed breathing configuration using the same force field.

Figure~\ref{fig:theory}(b-e) illustrates the theory predictions. In panel (b,c), the continuum model produces four imaginary branches and real-space mode patterns of \(1.08^\circ\) structure, matching the atomistic unstable modes of Fig.~\ref{fig:atomic}(a).  At angle \(4.41^\circ\), the same model predicts stable modes. Across the angle sweep, \(\eta\) decreases drastically. It starts from 5.27 at \(1.08^\circ\) to 1.04 at \(3.15^\circ\), 0.52 at \(4.41^\circ\), and 0.16 at \(7.34^\circ\), as in Fig.~\ref{fig:theory}(d). The dominant variation comes from the geometric factor $g^{-2}$, despite that the stress and bending related factor: \(S_{\rm eff}/\kappa_{\rm eff}\) (blue curve) varies much more gently (0.015 to 0.025). This isolates the dominant factor: the rapid increase of $g^{-2}$, expanded by over 45.8 times during the angle sweep. Therefore, the change of the overall $\eta$ binds strongly to the moir\'e length scale, which amplifies the rather smooth bending-stress balance. Fig.~\ref{fig:theory}(e) directly shows the effect of such a scale-driven phenomenon. The increase of imaginary mode numbers, the reduction of the lowest modes' eigenvalues, and the increase of the moir\'e wavelength occur simultaneously.

Fig.~\ref{fig:theory}(f-g) provides a mechanical aspect of the role played by the moir\'e length scale. The monotonic decrease of $A_1^{(1)}$ mode's AA area toward small twist angle shows it becomes increasingly confined to the AA-centred stress pocket. This confinement concentrates the local stress-induced negative stiffness, overcoming the bending penalty, making the flexural mode easier to destabilize at small angles while leaving the larger-angle structures stable. Together, these analyses explain mechanically the length scale driven MPC physics.

\begin{figure}[t!]
\centering
\includegraphics[width=\columnwidth]{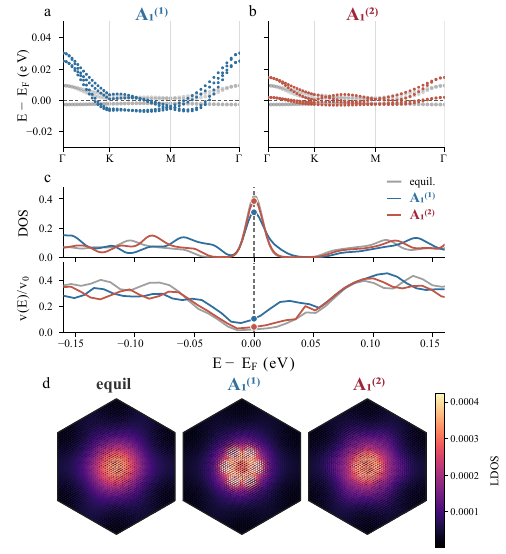}
\caption{\label{fig:electronic}
Electronic response to condensed moir\'e phonons. The band panels (a,b) compare result computed from the breathing reference (gray), with the structures displaced along $\mathbf{A}_1^{(1)}$ (blue) and $\mathbf{A}_1^{(2)}$ (red) mode. The energies are measured from the half-filled flat-band Fermi level. The DOS and diffusion-derived Fermi velocity in (c) shows $\mathbf{A}_1^{(1)}$ broadens the flat-band peak and increases metallic dispersion, whereas $\mathbf{A}_1^{(2)}$ leaves the bandwidth nearly intact. In panel (d), the LDOS is plotted at the Fermi-level, showing the clear electron redistribution by the $\mathbf{A}_1^{(1)}$ mode, forming a six-fold hexagon shape around the AA-region, while under the $\mathbf{A}_1^{(2)}$ mode it is only slightly renormalised.}
\end{figure}

We finally examine how the condensed moir\'e phonon coordinates affect the magic-angle flat electronic bands. Atomic corrugation and relaxation are known to modify magic-angle bandwidths, remote-band gaps, spectral weight, and local electronic reconstruction \cite{Yoo2019,Lucignano2019,Cantele2020,Rakib2022,Li2024}. Continuum and atomistic theories also show that lattice deformation enters the electronic Hamiltonian through interlayer tunnel coupling, orbital orientation, and moir\'e electron-phonon vertices \cite{KoshinoNam2020,Angeli2019,XieLiu2023,GuineaWalet2019,KangVafek2023,ZhuDevereaux2026}. The normal coordinates of the MPC allow us to resolve the structural-electronic coupling mode by mode.

We construct a $p_z$-orbital Slater-Koster tight-binding Hamiltonian on structures generated by freezing each active phonon coordinate separately,
\begin{equation}
    \mathbf{R}(Q_i)=\mathbf{R}_{\text{breathing}}+Q_i\mathbf{e}_i.
\end{equation}
Here $i=1$ or $2$. We use $Q_i$ and $e_i$ that derive from the REBO+KC force field. We then compare the resulting flat bands, density of states, Fermi velocity, and LDOS projected in real space. The Fermi velocity is extracted from the diffusion coefficient~\cite{guerrero2025disorder}, and is computed from a 16x16 supercell ($\sim$3M atoms) with a large-scale tight-binding method developed in-house. Details about the model parameters and the calculation can be found in Section IV of the supplemental materials.

Fig.~\ref{fig:electronic} shows that the two \(A_1\) coordinates have very different electronic fingerprints. Freezing $\mathbf{A}_1^{(1)}$ substantially increases the flat-band width, from 12.1 meV to 37.2 meV, and closes the \(2.7\mathrm{meV}\) gap inside the flat-band manifold. The same kinetic response is visible in the large-supercell DOS and diffusion-derived Fermi velocity: the flat-band DOS peak broadens, the extracted Fermi velocity increases by nearly a factor of three in electron side, from about \(0.1\) to \(0.3\) in the units of $v_0$, whereas the hole-side response is much weaker. In real space, the Fermi-level LDOS is transformed from a compact AA-centered peak (left) into a six-fold hexagonal texture (middle) as in Fig.~\ref{fig:electronic}(d). Thus \(A_1^{(1)}\) acts as an electronically active condensed coordinate, simultaneously increasing flat-band dispersion and transferring spectral weight from the AA center into six symmetry-related folds.

By contrast, the second active coordinate, \(A_1^{(2)}\), is electronically much weaker. Its band dispersion and DOS remain close to the relaxed reference, and the Fermi-level LDOS is only mildly renormalized. The AA-centered localization is largely preserved, with no comparable transfer of spectral weight. The two condensed coordinates, therefore, couple to distinct electronic channels. One mode primarily controls the flat-band kinetic energy and the internal angular texture of the AA-localized LDOS, while the other produces only a weak electronic response.

The MPC should be testable experimentally by starting from stable structures before the breathing-to-bending crossover. Low-frequency Raman or Brillouin light scattering may track the lowest layer-symmetric $A_1$ moir\'e flexural excitation near the moir\'e $\Gamma$ point, and observe the dynamical softening described by the MPC: $\omega_{A_1}^2(\theta)$ should decrease with decreasing $\theta$ and extrapolate toward zero. After the crossover, the breathing-saddle mode is replaced by vibrations around the bending minimum, while the layer-symmetric height field is frozen at finite mode amplitude. The combined observation of stable-side softening and the onset of bending would distinguish MPC from a purely geometric relaxation picture.  The experimental test should be possible since nano-Raman spectroscopy has resolved moir\'e-scale lattice-dynamical reconstruction in low-angle twisted bilayer graphene~\cite{Gadelha2021}, and low-frequency Raman experiments have observed twist-activated interlayer out-of-plane modes in twisted bilayer graphene~\cite{he2013observation}.

In conclusion, we identify the breathing-to-bending crossover as a Moir\'e Phonon Condensation, which is driven dominantly by the rapidly increasing moir\'e wavelength that amplifies a rather smooth stress-stiffness balance. Lowering twist angle \(\theta\) drives layer-symmetric \(A_1\) flexural phonons through zero stiffness, and the structure condenses into this moir\'e-scale phonon branch. The striking point is not only that imaginary modes exist and are controllable, but that a distortion involving all 11164 atoms and Angstrom-scale displacements remains almost completely confined to a two-dimensional vibrational subspace. Similar low-dimensional collapse is also found in twisted hBN, suggesting that Moir\'e Phonon Condensation is a broader structure reconstruction mechanism in twist materials. The condensed moir\'e phonon coordinates reshapes flat-band localization, bandwidth, transport velocity, and real-space electronic distribution. Moir\'e Phonon Condensation may therefore provide a way to think about reconstruction across moir\'e materials: as an emergent soft structural order parameter that can be tuned, coupled to, selectively excited, and ultimately used to control correlated electronic states.

\begin{acknowledgments}
 H.G. appreciates financial support from the Natural Science and Engineering Council (NSERC) of Canada. Z.Z. thanks the Department of Physics, McGill University, for Dr. and Mrs. Milton Leong Fellowships in Science. This work benefits from the co-authors' RQMP membership https://doi.org/10.69777/309032. We thank the Digital Research Alliance of Canada for substantial computation facility allocation which made this work possible. X.H.K. gratefully acknowledges the financial support from the Shenzhen Science and Technology Innovation Commission under the Outstanding Youth Project (Grant No. RCYX20231211090126026), the National Natural Science Foundation of China (Grants No. 12474173, 52461160327), Department of Science and Technology of Guangdong Province (Grants No. 2021QN02L820). 
\end{acknowledgments}

\bibliography{references}

\end{document}